\documentclass[aps,prl,reprint,longbibliography]{revtex4-1}

\usepackage{amsmath,amsfonts,bm,graphicx,hyperref,verbatim,mathrsfs,units,enumerate}

\usepackage{color,colordvi}
\definecolor{blue}{rgb}{0,0,1}
\definecolor{grey}{rgb}{0.6,0.6,0.6}

\begin{document}

\title{Detailed Fluctuation Relation for Arbitrary Measurement and Feedback Schemes}

\author{Patrick P. Potts}
\email{patrick.hofer@teorfys.lu.se}
\thanks{This author was previously known as Patrick P. Hofer.}
\author{Peter Samuelsson}
\affiliation{Physics Department and NanoLund, Lund University, Box 118,  22100 Lund, Sweden.}

\date{\today}

\begin{abstract}
Fluctuation relations are powerful equalities that hold far from equilibrium. However, the standard  approach to include measurement and feedback schemes may become inapplicable in certain situations, including continuous measurements, precise measurements of continuous variables, and feedback induced irreversibility. Here we overcome these shortcomings by providing a recipe for producing detailed fluctuation relations. Based on this recipe, we derive a fluctuation relation which holds for arbitrary measurement and feedback control. The key insight is that fluctuations inferable from the measurement outcomes may be suppressed by post-selection. Our detailed fluctuation relation results in a stringent and experimentally accessible inequality on the extractable work, which is saturated when the full entropy production is inferable from the data. 
\end{abstract}


\maketitle



\textit{Introduction.---}
Most devices that simplify our daily lives are far from equilibrium, consuming and dissipating energy. A thorough understanding of non-equilibrium physics is therefore of pivotal importance for the development of novel technologies. However, systems that are far from equilibrium are notoriously difficult to describe. This holds especially true for small systems, where fluctuations cannot be neglected. During the last $25$ years, a number of powerful thermodynamic equalities that hold far from equilibrium have been developed (for recent reviews, see Ref.~\cite{harris:2007,esposito:2009rmp,jarzynski:2011,seifert:2012,mansour:2017,campisi:2011,bochkov:2013}). The most prominent of these are the Jarzynski relation \cite{jarzynski:1997,jarzynski:1997pre} and the Crooks fluctuation theorem \cite{crooks:1998,crooks:1999,crooks:2000,kurchan:2000,tasaki:2000} (see also Refs.~\cite{bochkov:1981,bochkov:1981b}). These equalities involve the probability distributions of work or entropy production along trajectories through phase space and constitute important results in the field of stochastic thermodynamics.

Recent experimental advances in observing and controlling small systems opened up the possibility of optimizing the process at hand using feedback control \cite{ciliberto:2017}. Promising platforms for such experiments include electronic systems \cite{koski:2014,koski:2015,hofmann:2016,hofmann:2017,chida:2017}, DNA molecules \cite{alemany:2012,dieterich:2016}, photons \cite{vidrighin:2016}, Brownian particles \cite{toyabe:2010}, and superconducting circuits in the quantum regime \cite{cottet:2017,masuyama:2018,naghiloo:2018}. These experiments probe the thermodynamics of information \cite{maxwell:book,sagawa:prog,maruyama:2009,parrondo:2015}, a field which goes back to the thought experiments of Maxwell and Szilard \cite{maxwell:1871,szilard:1929,rex:2017}, where microscopic information is used to seemingly violate the second law and to produce useful work. 
Under measurement and feedback schemes, fluctuation relations and second-law-like inequalities can still be derived by including a term that represents the obtained information \cite{sagawa:2008,cao:2009,sagawa:2010,horowitz:2010,ponmurugan:2010,morikuni:2011,sagawa:2012,sagawa:2012prl,sagawa:2013,lahiri:2012,abreu:2012,funo:2013,ashida:2014,horowitz:2014,horowitz:2014prx,funo:2015,wachtler:2016,gong:2016,spinney:2016,spinney:2018,kwon:2017}. 
For the Jarzynski relation, the most prominent generalizations read \cite{sagawa:2010,sagawa:2012}
\begin{align}
\label{eq:jarz1}
\left\langle e^{-\sigma-I}\right\rangle&=1\hspace{.25cm}\Rightarrow\hspace{.25cm} \langle \sigma\rangle\geq -\langle I\rangle,\\
\label{eq:jarz2}
\left\langle e^{-\sigma}\right\rangle&=\gamma\hspace{.25cm}\Rightarrow\hspace{.25cm} \langle \sigma\rangle\geq -\ln\gamma,
\end{align}
where $I$ denotes the transfer entropy (the average of which reduces to the mutual information for a single measurement), $\gamma$ the efficacy parameter, and $\sigma$ the entropy production. 

While existing fluctuation relations constitute powerful results, they are unfortunately not always applicable and a detailed fluctuation relation for arbitrary measurement and feedback scenarios is still lacking. The problems that can arise can be exemplified with the help of Eqs.~\eqref{eq:jarz1} and \eqref{eq:jarz2}, where we identified three key shortcomings:
(i) The quantities $I$, $\langle I\rangle$, and $\gamma$ can diverge, rendering Eqs.~\eqref{eq:jarz1} and \eqref{eq:jarz2} inapplicable. In particular, $I$ diverges when the feedback introduces absolute irreversibility. A naive evaluation of the Jarzynski relation in Eq.~\eqref{eq:jarz1} then yields the wrong result \cite{horowitz:2010,murashita:2014}. The average of the transfer entropy $\langle I\rangle$ can diverge, e.g., for continuous measurements, when the amount of information extracted from the system diverges \cite{horowitz:2014}. Moreover, the efficacy parameter $\gamma$ can diverge for feedback schemes that include a large number of control protocols to choose from (see below).
(ii) The transfer entropy $I$ is not directly measurable as it contains information on the correlations between system and measurement apparatus \cite{sagawa:2012prl,sagawa:2013}. This limits the practical relevance of Eq.~\eqref{eq:jarz1}. 
(iii) For Eq.~\eqref{eq:jarz2}, there is to date no corresponding detailed fluctuation relation which relates probabilities in a \textit{forward} experiment to probabilities in a \textit{backward} experiment.
Given these shortcomings, it is highly desirable to obtain refined detailed fluctuation relations which hold for any measurement and feedback scheme. For error-free measurements, an effort in this direction has been made in Ref.~\cite{ashida:2014}.

In this Letter, we overcome the shortcomings of fluctuation relations in the presence of measurement and feedback with two interrelated contributions. First, we provide a novel recipe for obtaining fluctuation relations. Upon defining a backward experiment our recipe provides the associated fluctuation relation, including the corresponding information terms. This allows one to tailor useful fluctuation relations, Jarzynski relations, and second-law-like inequalities for the problem at hand. Second, we use this recipe to find a detailed fluctuation relation that circumvents the problems (i)-(iii) listed above. 
In the case of error-free measurements, our fluctuation relation reduces to the one found in Ref.~\cite{ashida:2014}.

\begin{figure}
\centering
\includegraphics[width=\columnwidth]{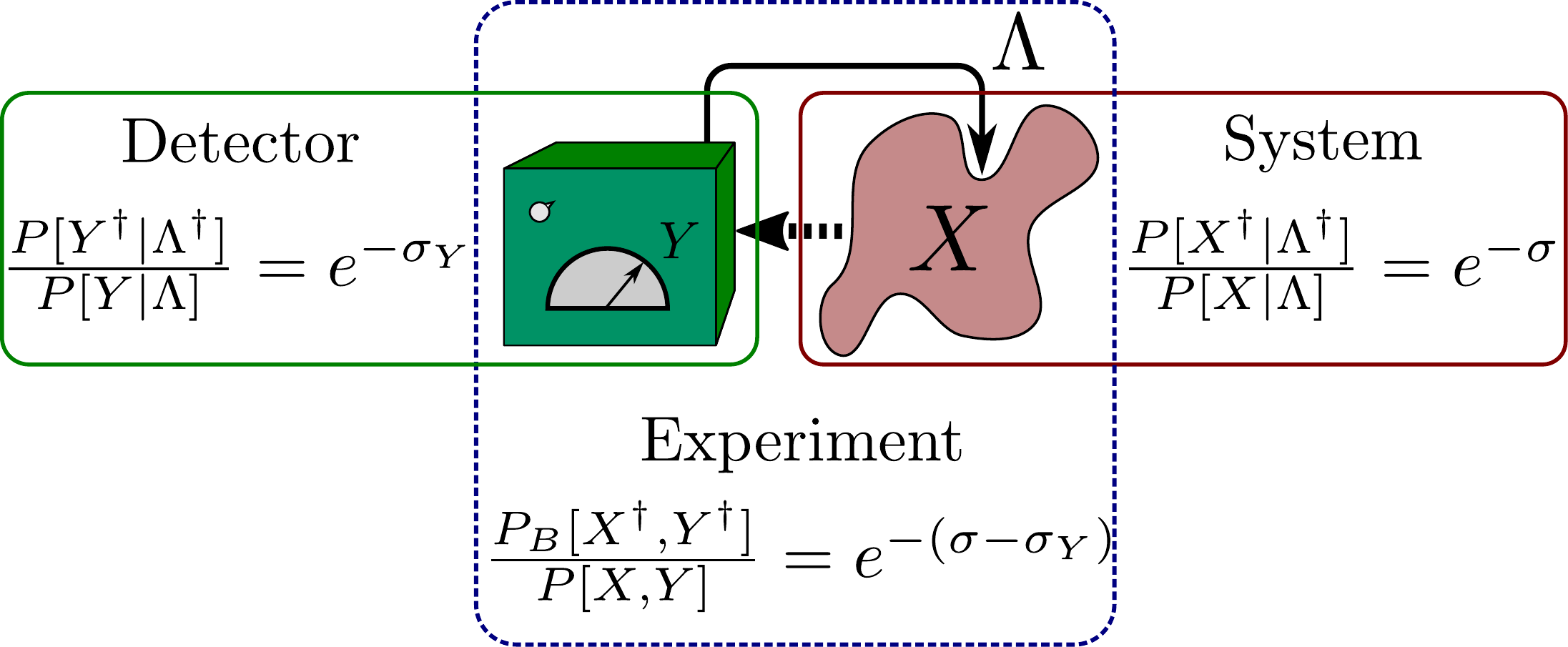}
\caption{Illustration of the fluctuation relation for measurement and feedback. Both the system as well as the detector output fulfill a detailed fluctuation relation. Here $X$ ($Y$) denotes a trajectory of the system state (detector output) and $\Lambda$ {a trajectory of the control parameter}. A detailed fluctuation relation for the full experiment can be obtained, where the total entropy production $\sigma$ is reduced by the inferable entropy production $\sigma_Y$.
	 Probability distributions are defined in the text.}
  \label{fig:model}
\end{figure}

\textit{A recipe for fluctuation relations.---} Our starting point is the detailed fluctuation relation for a fixed control protocol, a fundamental relation which generalizes the second law for stochastic systems \cite{gallavotti:1995prl,gallavotti:1995,kurchan:1998,crooks:1998,crooks:1999,crooks:2000,jarzynski:2000,seifert:2005,sagawa:2012}. In the notation of Ref.~\cite{sagawa:2012}, largely followed throughout this Letter, we have
{\begin{equation}
\label{eq:detailflucsys}
\frac{P[X^\dagger|\Lambda^\dagger]}{P[X|\Lambda]}=e^{-\sigma[X,\Lambda]}.
\end{equation}}
Here the vector $X=(x_1,\cdots,x_N)$ denotes a system trajectory through phase space, where time is discretized and $x_j$ denotes the point in phase space the system occupies at time $t_j$. The time-step $t_{j+1}-t_j=\delta t$ is assumed to be infinitesimally small. {Similarly, $\Lambda=(\lambda_1,\cdots,\lambda_N)$ denotes a trajectory of the control parameter (sometimes called protocol). For instance, $\lambda_j$ can be the value of an electric field at time $t_j$.} The daggered quantities denote the time-reverse of the undaggered ones, e.g., $X^\dagger=(x_N^*,\cdots,x_1^*)$, where $x_j^*$ is the time-reverse of $x_j$ {and similarly for $\Lambda$. Note that the daggered quantities are uniquely defined by the undaggered ones.}

Equation \eqref{eq:detailflucsys} can be understood as follows: $P[X|\Lambda]$ denotes the probability that the system takes trajectory $X$ when {the control parameter is determined by $\Lambda$}. The probability $P[X^\dagger|\Lambda^\dagger]$ of realizing the time-reversed trajectory when applying the time-reversed control parameter is related to $P[X|\Lambda]$ by the exponentiated entropy production \cite{seifert:2005} (see the supplemental information below for a general definition).
For experiments that start in thermal equilibrium, and systems coupled to a single bath at temperature $T$, the entropy production can be written as
\begin{equation}
\label{eq:entropy}
k_B T \sigma[X,\Lambda]=\Delta F[\Lambda]-W[X,\Lambda],
\end{equation}
where $\Delta F[\Lambda]$ corresponds to the free energy difference of the equilibrium states at the beginning and at the end of the experimental run and $W[X,\Lambda]$ denotes the work \textit{extracted} from the system.

{To include measurement and feedback, we denote by $Y=(y_1,\cdots,y_N)$ a trajectory of measurement outcomes, encoding information on $X$. Discrete measurements can be obtained by taking most $y_j$ independent of the system trajectory. Feedback is included by determining the control parameter based on the measurement outcomes, i.e. $\Lambda(Y)$. We stress that Eq.~\eqref{eq:detailflucsys} is still valid since it only involves probabilities which are conditioned on the control parameter. For ease of notation, we omit the $Y$-dependence of $\Lambda$ whenever there is no explicit $Y$-dependence.}


In the presence of measurement and feedback, the forward experiment is described by a joint probability distribution for system trajectory $X$ and measurement outcome $Y$
 \cite{sagawa:2012}
 \begin{equation}
 \label{eq:probaforward}
 P[X,Y]=P_m[Y|X]P[X|\Lambda(Y)],
 \end{equation}
 where $P_m[Y|X]$ denotes the probability that a fixed trajectory $X$ results in the measurement outcomes $Y$. {For more details, see Ref.~\cite{sagawa:2012}. For our purposes, the last equation can be seen as the definition of $P_m[Y|X]$. Equation \eqref{eq:probaforward} illustrates that a feedback experiment includes two ingredients. 1. A set of possible trajectories for the control parameter, and 2. a decision procedure to determine which trajectory is applied. Throughout this Letter, an experiment is defined by these two ingredients as well as a possible third one: 3. post-processing of the measured data.
 	
 In the absence of measurement and feedback, there is usually only a single trajectory for the control parameter and ingredients 2 and 3 are unnecessary. The detailed fluctuation relation in Eq.~\eqref{eq:detailflucsys} then relates the forward experiment to the backward experiment, which is provided by applying the time-reverse of the control parameter trajectory. In the presence of measurement and feedback, defining a backward experiment is much less trivial. While the control parameter trajectories can simply be time-reversed, it is not a priori clear how to fix ingredients 2 and 3 As we will now discuss in detail, this freedom in choosing the backward experiment results in many different fluctuation relations.}

We note that if not specifically stated otherwise, our results only require Eq.~\eqref{eq:detailflucsys} to hold and do not depend on the specifics of the entropy production. For non-equilibrium initial states, there are cases when Eq.~\eqref{eq:detailflucsys} becomes inapplicable \cite{murashita:2014,funo:2015}. This problem can be circumvented by including the preparation of the initial state in the process.
 	
Rewriting Eq.~\eqref{eq:detailflucsys}, we arrive at our first main contribution, a general detailed fluctuation relation for joint probabilities
\begin{equation}
\label{eq:fluctheoremgen}
\frac{P_B[X^\dagger,Y^\dagger]}{P[X,Y]}=e^{-\sigma[X,\Lambda(Y)]-(I[X:Y]-I^\dagger[X^\dagger:Y^\dagger])},
\end{equation}
where $P_B[X^\dagger,Y^\dagger]$ denotes the probability distribution for the backward experiment; unspecified thus far. Here we introduced the transfer entropy in the forward experiment
\begin{equation}
\label{eq:ic}
I[X:Y]=\ln\frac{P[X,Y]}{P[X|\Lambda(Y)]P[Y]}=\ln\frac{P_m[Y|X]}{P[Y]},
\end{equation}
and in the backward experiment
\begin{equation}
\label{eq:icdag}
I^\dagger[X^\dagger:Y^\dagger]=\ln\frac{P_B[X^\dagger,Y^\dagger]}{P[X^\dagger|\Lambda(Y)^\dagger]P[Y]},
\end{equation}
and $P[Y]=\int dXP[X,Y]$. To illustrate the usefulness of Eq.~\eqref{eq:fluctheoremgen} as a recipe for fluctuation relations, we consider the following scenario: An experiment using measurement and feedback has been designed and it is desired to investigate the physics of the experiment with fluctuation relations. While the forward experiment is fixed by the designed experiment, there is a freedom in choosing the backward experiment. For any chosen backward experiment, Eq.~\eqref{eq:fluctheoremgen} provides a fluctuation relation and allows for identifying the corresponding information terms.

It is instructive to see how previous results can be recovered from Eq.~\eqref{eq:fluctheoremgen}. To this end, we consider a backward experiment where no feedback is performed. Instead, the fixed {control parameter} $\Lambda^\dagger$ is performed with the same probability as $\Lambda$ is applied in the forward experiment (where it arises from feedback). This corresponds to the backward probability $P_B[X^\dagger,Y^\dagger]=P[X^\dagger|\Lambda(Y)^\dagger]P[Y]$. Equation \eqref{eq:fluctheoremgen} then results in the fluctuation relation associated to Eq.~\eqref{eq:jarz1} \cite{sagawa:2010,sagawa:2012}. 
Here we mainly focus on scenarios where $P_B$ describes an actual experiment and is thus a normalized probability distribution. However, for any function $P_B$, Eq.~\eqref{eq:fluctheoremgen} can be used to derive integral fluctuation relations. For instance, we can recover the integral fluctuation relation in Eq.~\eqref{eq:jarz2} by choosing $P_B[X^\dagger,Y^\dagger]=P_m[Y|X]P[X^\dagger|\Lambda(Y)^\dagger]$, which is not a normalized probability distribution. {Indeed, when Eq.~\eqref{eq:trs} below holds, this distribution is normalized to the efficacy parameter $\gamma$.}

{Other definitions of $P_B$ will result in different fluctuation relations.}
More generally, one can demand conditions on the backward experiment and/or the information terms in Eq.~\eqref{eq:fluctheoremgen} to find novel fluctuation relations. Generalized Jarzynski relations and second-law-like inequalities can then be derived in a straightforward manner.

\textit{A versatile fluctuation relation.---} We now apply our recipe to find a fluctuation relation which circumvents the shortcomings (i)-(iii) listed in the introduction. To this end, we impose two conditions:
\begin{enumerate}[I]
\item \label{cond1} The quantity $\Delta I[Y]\equiv I[X:Y]-I^\dagger[X^\dagger:Y^\dagger]$ {shall be} fully determined by the measurement outcomes.
\item \label{cond2} The $Y$-marginals of the forward and backward probabilities {shall be} the same $\int dX P_B[X^\dagger,Y^\dagger]=P[Y]$.
\end{enumerate}
The first condition ensures that the information term $\Delta I$ is experimentally accessible, overcoming shortcoming (ii). The second condition demands that a given set of measurement outcomes $Y$ is equally likely in the forward and in the backward experiment.

These two conditions {uniquely fix $P_B$ in} Eq.~\eqref{eq:fluctheoremgen}, resulting in our second main contribution, a detailed fluctuation relation applicable for arbitrary measurement and feedback scenarios.
{We now discuss both the backward probability distribution as well as the information term derived from our conditions (see the supplemental information for detailed derivations). First, we have $\Delta I [Y]=-\sigma_{\rm cg}$,}
where we introduced the coarse-grained entropy production \cite{kawai:2007,sagawa:2012}
\begin{equation}
\label{eq:entcg}
e^{-\sigma_{\rm cg}[Y]}\equiv\int dX e^{-\sigma[X,\Lambda(Y)]}P[X|Y],
\end{equation}
where $P[X|Y]=P[X,Y]/P[Y]$. We note that as long as the total entropy production remains finite, $\sigma_{\rm cg}$ remains finite as well, preventing the divergences related to shortcoming (i).
{We find a generalized Jarzynski relation including the coarse-grained entropy production}
\begin{equation}
\label{eq:jarent}
\left\langle e^{-(\sigma-\sigma_{\rm cg}[Y])} \right\rangle = 1\hspace{.25cm}\Rightarrow\hspace{.25cm}\langle\sigma\rangle \geq  \langle \sigma_{\rm cg}[Y]\rangle,
\end{equation}
where $\langle\cdots\rangle$ denotes an average over the forward probability distribution and the second-law-like inequality follows from Jensen's inequality.

\begin{figure*}
	\centering
	\includegraphics[width=\textwidth]{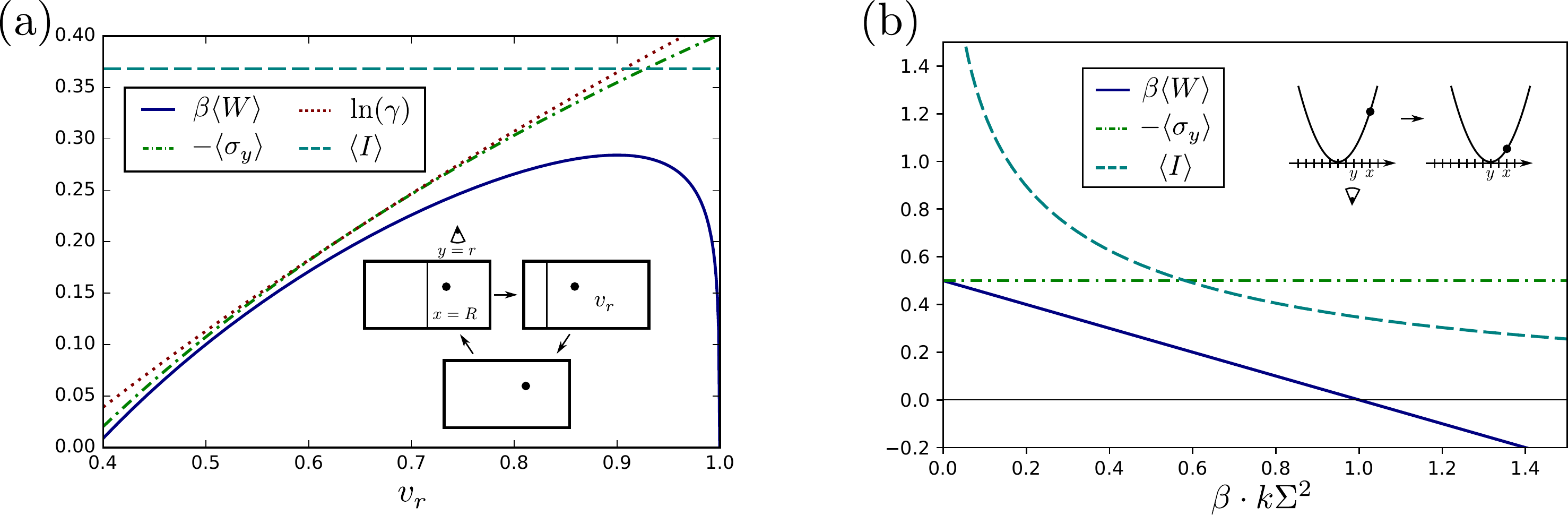}
	\caption{Second-law-like bounds for the extracted work. The extracted work (blue, solid) is compared to the inferable entropy production (green, dash-dotted), the logarithm of the efficacy parameter (red, dotted), and the transfer entropy (cyan, dashed). (a) Szilard engine. On the horizontal axis, the final volume for measurement outcome $y=r$ is varied. For a broad range of parameters, the inferable entropy provides the tightest bound on the extracted work. Here the measurement error probability is $\varepsilon=0.1$ and the final volume for measurement outcome $y=l$ is $v_l=0.65$. (b) Brownian particle in a harmonic trap. On the horizontal axis, the measurement error $\Sigma^2$ divided by $k_BT/k$ is varied, where $k$ denotes the spring constant of the trap. The transfer entropy diverges as the measurement error goes to zero and the efficacy parameter diverges for all parameters. The inferable entropy provides a bound that becomes tighter as the measurement becomes more precise. We note that in both examples, the transfer entropy equals the mutual information since there is only a single measurement.}
	\label{fig:examples}
\end{figure*}

Of key importance are scenarios which fulfill the measurement time-reversal symmetry
\begin{equation}
\label{eq:trs}
P_m[Y|X]=P_m[Y^\dagger|X^\dagger].
\end{equation} 
As we will see below, this condition leads to a particularly illuminating physical interpretation of our fluctuation relation and ensures that the backward probability distribution has an operational meaning. We also note that this condition underlies Eq.~\eqref{eq:jarz2}.
Given Eq.~\eqref{eq:trs}, it can be shown that a detailed fluctuation relation for the detector output holds \cite{sagawa:2012}
\begin{equation}
\label{eq:detailflucdet}
e^{-\sigma_{\rm cg}[Y]}=e^{-\sigma_Y}\equiv\frac{P[Y^\dagger|\Lambda(Y)^\dagger]}{P[Y|\Lambda(Y)]},
\end{equation}
where $P[Y|\Lambda]=\int dXP_m[Y|X]P[X|\Lambda]$ denotes the probability of obtaining the outcomes $Y$ given the control parameter $\Lambda$. From Eq.~\eqref{eq:probaforward}, we thus find $P[Y|\Lambda(Y)]=P[Y]$. Comparing Eq.~\eqref{eq:detailflucdet} with the detailed fluctuation relation in Eq.~\eqref{eq:detailflucsys}, we conclude that $\sigma_Y$ is the entropy production that we infer from observing only the measurement outcomes (see also Fig.~\ref{fig:model}). We thus call it the \textit{inferable entropy production}. We note that the coarse-grained entropy production is only equal to the inferable entropy production when Eq.~\eqref{eq:trs} holds. In the following, we thus identify $\sigma_Y=\sigma_{\rm cg}$, deferring a discussion on scenarios where this is not the case to the supplemental information.
Equation \eqref{eq:detailflucdet} implies $\langle \exp{(-\sigma_Y)}\rangle =\gamma$. From Jensen's inequality we then find $\langle \sigma_Y\rangle\geq -\ln\gamma$. The inequality in Eq.~\eqref{eq:jarent} is thus strictly more stringent than the inequality based on the efficacy parameter given in Eq.~\eqref{eq:jarz2}.

The backward probability obtained from our conditions I, II, and Eq.~\eqref{eq:trs} reads
\begin{equation}
\label{eq:pback}
P_B[X^\dagger,Y^\dagger]=\frac{P[X^\dagger|\Lambda(Y)^\dagger]}{P[Y^\dagger|\Lambda(Y)^\dagger]}P_m[Y^\dagger|X^\dagger]P[Y].
\end{equation}
This distribution has an operational meaning [overcoming shortcoming (iii)] and can be obtained as follows: In a backward experiment, the {control parameter} $\Lambda(Y)^\dagger$ is applied with probability $P[Y]$. {Just as in Ref.~\cite{sagawa:2012}, $Y^\dagger$ is thus determined probabilistically at the beginning of each experimental run.} {The same measurements as in the forward experiment are then carried out but in time-reversed order. Importantly, the measurement outcomes are not used to update the control parameter. The data is then post-selected, discarding all experimental runs where the measurement outcomes are not equal to $Y^\dagger$ when applying $\Lambda(Y)^\dagger$. The distribution $P_B[X^\dagger,Y^\dagger]$ is the joint probability for realizing $X^\dagger$ and $Y^\dagger$ in this backward experiment.} It is the post-selection which results in the reduction of the entropy production by the inferable entropy production $\sigma_Y$. Intuitively, having access to the measurement outcomes, their fluctuations can be suppressed. This is illustrated in Fig.~\ref{fig:model}.
In case the full entropy production is inferable from the measurement outcomes, i.e., $\sigma_Y=\sigma$, our fluctuation relation reduces to the trivial equality $1=1$ reflecting the fact that the full entropy production is accessible. Finding deviations from this trivial identity then reflects the fact that not all entropy producing degrees of freedom are perfectly measured. To verify this, the entropy production must be measurable independently from $Y$.

{Under our conditions, one can integrate Eq.~\eqref{eq:fluctheoremgen} over all $X$ which result in the same $\sigma$ to obtain a fluctuation relation for entropy production (see supplemental information). We note that this is not generally possible for previous fluctuation relations.}
For an entropy production given by Eq.~\eqref{eq:entropy}, this results in a fluctuation relation for the extracted work $W$
\begin{align}
\label{eq:flucwork}
\frac{P[W,Y]}{P_B[-W,Y^\dagger]}&=e^{-\beta (W-\Delta F[\Lambda(Y)])-\sigma_Y},\\
\label{eq:jenwork}
\Rightarrow\langle W\rangle& \leq  \langle\Delta F[\Lambda(Y)]\rangle-k_B T \langle \sigma_Y\rangle,
\end{align}
{where $P[W,Y]$ is the joint probability of obtaining a value $W$ for the work and a measurement outcome equal to $Y$ in the forward experiment (and similarly for the backward experiment).}
We note that in the absence of feedback, the probability distributions factorize and Eq.~\eqref{eq:flucwork} reduces to a simple product between the Crooks fluctuation relation and Eq.~\eqref{eq:detailflucdet}.
To illustrate our results, we consider two well-studied examples, the Szilard engine and a Brownian particle in a harmonic trap. We note that Eq.~\eqref{eq:trs} holds for both examples.

\textit{The Szilard engine.---} We consider a particle in a box of volume $v=1$. A separation in the middle of the box is introduced and the particle will be found to the left $x=L$ or to the right $x=R$ of the separation with equal probabilities. Subsequently, the location of the particle is measured with an error $\varepsilon$ resulting in a measurement outcome $y\in\{l,r\}$. The separation is then slowly moved with the aim of increasing the volume available to the particle to $v_y$, depending on the outcome of the measurement. Finally, the separation is removed and the system returns to its initial state.

Detailed calculations are given in the supplemental information, where we verify the detailed fluctuation relation given in Eq.~\eqref{eq:flucwork}.
In Fig.~\ref{fig:examples}\,(a), we show the extracted work and compare it to the bounds given in Eqs.~\eqref{eq:jarz1}, \eqref{eq:jarz2}, and \eqref{eq:jenwork}. We find that the inequality involving the inferable entropy production gives a tighter bound than the established inequalities for a range of parameters.

\textit{Brownian particle in a harmonic trap.---} Our second example consists of a Brownian particle in a harmonic trap potential with spring constant $k$. After a position measurement is performed, the trap potential is shifted, such that the new minimum coincides with the measurement outcome. As long as the thermal spread, $k_BT/k$ is larger than the measurement error, denoted by $\Sigma^2$, a positive amount of work is extracted from the particle on average. As for the Szilard engine, detailed calculations are given in the supplemental information where Eq.~\eqref{eq:flucwork} is explicitly verified. In Fig.~\ref{fig:examples}\,(b), the extracted work is compared to the transfer entropy and the inferable entropy production. The efficacy parameter diverges in this scenario since the position measurement has infinitely many outcomes, resulting in infinitely many {control parameter trajectories}. The transfer entropy diverges as the measurement error goes to zero. The inferable entropy production provides a useful bound for all parameters. We note that Ref.~\cite{ashida:2014} discussed the same example in the limit $\Sigma\rightarrow 0$, where the bound provided by the inferable entropy becomes tight.

As an additional example published elsewhere, our results are applied to continuous measurements in single molecule force spectroscopy experiments \cite{schmitt:prep}.

\textit{Conclusions.---} We provided a recipe for obtaining fluctuation relations in the presence of measurement and feedback. This recipe relies on the freedom of choosing a backward experiment and can be employed to develop useful and experimentally relevant fluctuation relations. This is illustrated with a detailed fluctuation relation which overcomes the shortcomings identified in previous works. The resulting relation allows for an intuitive explanation and provides a second-law like inequality in situations where previous fluctuation relations break down.


The freedom of choosing a backward experiment indicates that there is no single fluctuation relation which is universally optimal, but that each class of problems might be best described by a tailor-made fluctuation relation. The general validity of our recipe allows for the construction of relevant fluctuation relations for any given problem including measurement and feedback. The approach outlined here has thus great potential for obtaining a better understanding of non-equilibrium processes and will likely result in additional practically useful equalities and inequalities.

\textit{Acknowledgements.---} We acknowledge insightful comments by M. Ueda and F. Ritort as well as fruitful discussions with R. K. Schmitt and C. Van den Broeck. This work was supported by the Swedish Research Council. P.P.P. acknowledges funding from the European Union's Horizon 2020 research and innovation programme under the Marie Sk{\l}odowska-Curie Grant Agreement No. 796700.

\bibliography{biblio}

\clearpage
\widetext

\begin{center}
	\textbf{\large Supplemental information: A Detailed Fluctuation Relation for Arbitrary Measurement and Feedback Schemes}
\end{center}
\setcounter{equation}{0}
\setcounter{figure}{0}
\setcounter{table}{0}
\setcounter{page}{1}
\makeatletter
\renewcommand{\theequation}{S\arabic{equation}}
\renewcommand{\thefigure}{S\arabic{figure}}
This supplemental information provides the general definition of the entropy production, a summary of the employed probability distributions in the main text, a derivation of the fluctuation theorem discussed in the main text, a discussion on measurements without time-reversal symmetry, as well as detailed derivations for the examples in the main text. Equation and Figure numbers not preceded by an `$S$' refer to the main text.

\section{A. Entropy production}
The general definition of the entropy production that enters Eq.~\eqref{eq:detailflucsys} in the main text reads
\begin{equation}
\label{eq:entropys}
\sigma[X,\Lambda]=\ln p(x_1|\lambda_1)-\ln p(x_N^*|\lambda_N^*)-\sum_{\alpha}\beta_\alpha Q_\alpha[X,\Lambda].
\end{equation}
Here $p(x_1|\lambda_1)$ and $p(x_N^*|\lambda_N^*)$ denote the initial distribution for the forward and backward experiment respectively. The system is further assumed to be coupled to thermal baths labeled by the index $\alpha$ which are at the inverse temperature $\beta_\alpha$. The heat which enters the system from bath $\alpha$ is denoted by $Q_\alpha$. We note that the initial distributions can in principle take any shape. However, we focus on experiments that start in thermal equilibrium and on systems coupled to a single bath at temperature $T$. In this case, the entropy production reduces to Eq.~\eqref{eq:entropy} in the main text
\begin{equation}
\label{eq:entropy2}
k_B T \sigma[X,\Lambda]=\Delta F[\Lambda]-\Delta E[X,\Lambda]- Q[X,\Lambda]=\Delta F[\Lambda]-W[X,\Lambda],
\end{equation}
where the difference in the system energy is given by $\Delta E[X,\Lambda]=E(x_1,\lambda_1)-E(x_N,\lambda_N)$, and the difference in the free energy is given by $\Delta F[\Lambda]=F(\lambda_1)-F(\lambda_N)$. Here we assumed the symmetry $E(x,\lambda)=E(x^*,\lambda^*)$. For the second equality in Eq.~\eqref{eq:entropy2}, we used the first law of thermodynamics $\Delta E = W-Q$, where $W$ is the work \textit{extracted} from the system.

\section{B. Probability distributions employed in the main text}
The central probability distribution in the main text is given by (see Ref.~\cite{sagawa:2012} for a detailed derivation)
\begin{equation}
\label{eq:jointprobafeedback}
P[X,Y]=P_m[Y|X]P[X|\Lambda(Y)].
\end{equation}
This distribution gives the joint probability that the system follows trajectory $X$ and the measurement outcomes are given by $Y$. The last equation tells us that the joint probability distribution in a feedback experiment is given by the product of two distributions. $P[X|\Lambda]$ denotes the probability that the system takes trajectory $X$ when the control parameter follows the fixed trajectory $\Lambda$. $P_m[Y|X]$ denotes the probability that the measurement outcomes are $Y$ if the system trajectory is fixed to be $X$. 

In addition to these distributions, Bayes theorem is applied in the main text
\begin{equation}
\label{eq:bayessupp}
P[X,Y]=P[X|Y]P[Y]=P[Y|X]P[X],
\end{equation}
where the marginal probability distributions read
\begin{equation}
\label{eq:marginalssupp}
P[X] = \int dY P[X,Y],\hspace{.2cm}P[Y] = \int dY P[X,Y].
\end{equation}
Note that the conditional probability distribution $P[Y|X]\neq P_m[Y|X]$. The reason for this is that $P_m[Y|X]$ is not just conditioned on $X$ but on $X$ given the protocol $\Lambda(Y)$ [cf.~Eq.~\eqref{eq:jointprobafeedback}].

Finally, we make use of the distribution
\begin{equation}
\label{eq:ygivenlamsupp}
P[Y|\Lambda]=\int dXP_m[Y|X]P[X|\Lambda].
\end{equation}
This distribution gives the probability to observe the outcome $Y$ under the protocol $\Lambda$. Importantly, we can choose a protocol which is different from $\Lambda(Y)$ in the last expression. However, if inserting $\Lambda(Y)$ in the last expression, then we find $P[Y|\Lambda(Y)]=P[Y]$ from Eqs.~\eqref{eq:marginalssupp} and \eqref{eq:jointprobafeedback}.

\section{C. Derivation of the fluctuation theorem}
Using condition I in the main text, we can write the detailed fluctuation relation in Eq.~\eqref{eq:fluctheoremgen} as
\begin{equation}
\label{eq:fluctheoremgensupp}
\frac{P_B[X^\dagger,Y^\dagger]}{P[X,Y]}=e^{-\sigma[X,\Lambda(Y)]-\Delta I[Y]}.
\end{equation}	
From this follows
\begin{equation}
\label{eq:der10s2}
\begin{aligned}
\int dX P_B[X^\dagger,Y^\dagger] &= e^{-\Delta I[Y]}\int dX P[X,Y]e^{-\sigma[X,\Lambda(Y)]} \\&= P[Y]e^{-\Delta I[Y]}\int dX P[X|Y]e^{-\sigma[X,\Lambda(Y)]} =P[Y]e^{-\Delta I[Y]-\sigma_{\text{cg}}[Y]},
\end{aligned}
\end{equation}
where we used Bayes theorem and the definition of the coarse-grained entropy production in Eq.~\eqref{eq:entcg} From condition II, it then follows that $\Delta I[Y]=-\sigma_{\text{cg}}[Y]$ and we obtain the detailed fluctuation relation
\begin{equation}
\label{eq:ntrdetail1}
\frac{P_B[X^\dagger,Y^\dagger]}{P[X,Y]}=e^{-(\sigma[X,\Lambda(Y)]-\sigma_{\text{cg}}[Y])}.
\end{equation}	
To obtain a fluctuation relation for entropy production, we write
\begin{equation}
\label{eq:flucsigsupp}
\begin{aligned}
&\int dX \delta(\sigma[X,\Lambda(Y)]-\sigma) P_B[X^\dagger,Y^\dagger]= \int dX \delta(\sigma[X,\Lambda(Y)]-\sigma)P[X,Y]e^{-(\sigma[X,\Lambda(Y)]-\sigma_{\text{cg}}[Y])} \\\Rightarrow&\int dX \delta(\sigma[X^\dagger,\Lambda(Y)^\dagger]+\sigma) P_B[X^\dagger,Y^\dagger]=e^{-(\sigma-\sigma_{\text{cg}}[Y])}\int dX \delta(\sigma[X,\Lambda(Y)]-\sigma)P[X,Y],
\end{aligned}
\end{equation}
where we used $\sigma[X^\dagger,\Lambda(Y)^\dagger]=-\sigma[X,\Lambda(Y)]$. We now define
\begin{equation}
\label{eq:psigsup}
P[\sigma,Y]=\int dX \delta(\sigma[X,\Lambda(Y)]-\sigma)P[X,Y],
\end{equation}
as well as
\begin{equation}
\label{eq:psigdagsup}
P_B[\sigma,Y^\dagger]=\int dX \delta(\sigma[X,\Lambda(Y)^\dagger]-\sigma) P_B[X,Y^\dagger],
\end{equation}
where we used $dX=dX^\dagger$. {$P[\sigma,Y]$ is thus the joint probability distribution to have the entropy $\sigma$ and observe the measurement outcome $Y$ in a run of the forward experiment (and analogously for the backward experiment). We note that these probabilities can only be accessed experimentally if the entropy production can be measured. from these definitions, we find
	\begin{equation}
	\label{eq:fluctheorent}
	\frac{P_B[-\sigma,Y^\dagger]}{P[\sigma,Y]}=e^{-(\sigma-\sigma_{\text{cg}}[Y])}.
	\end{equation}
	When the entropy production is given by Eq.~\eqref{eq:entropy2}, an analogous calculation results in Eq.~\eqref{eq:flucwork} in the main text.}

To determine the backward probability distribution, we write
\begin{equation}
\label{eq:cgentsupp}
e^{-\sigma_{\rm cg}[Y]}=\int dX e^{-\sigma[X,\Lambda(Y)]}P[X|Y]=\int dX \frac{P[X^\dagger|\Lambda(Y)^\dagger]}{P[X|\Lambda(Y)]}\frac{P[X,Y]}{P[Y]}=\frac{\int dX P[X^\dagger|\Lambda(Y)^\dagger]P_m[Y|X] }{P[Y]},
\end{equation}
where we used Bayes theorem and Eq.~\eqref{eq:detailflucsys} in the main text in the first equality and Eq.~\eqref{eq:jointprobafeedback} in the second equality. Inserting Eq.~\eqref{eq:cgentsupp} and Eq.~\eqref{eq:detailflucsys} from the main text into Eq.~\eqref{eq:ntrdetail1}, we find
\begin{equation}
\label{eq:backwardprobantr1}
P_B[X^\dagger,Y^\dagger]=\frac{P_m[Y|X]P[X^\dagger|\Lambda(Y)^\dagger]}{\int dX P_m[Y|X]P[X^\dagger|\Lambda(Y)^\dagger]}P[Y].
\end{equation}
Under the assumption $P_m[Y^\dagger|X^\dagger]= P_m[Y|X]$, and using Eq.~\eqref{eq:ygivenlamsupp}, we recover Eqs.~\eqref{eq:detailflucdet} and \eqref{eq:pback} from Eqs.~\eqref{eq:cgentsupp} and \eqref{eq:backwardprobantr1} respectively.

\section{D. Measurements without time-reversal symmetry}
Here we consider the case where $P_m[Y^\dagger|X^\dagger]\neq P_m[Y|X]$ in some detail. In this case, the two conditions given in the main text result in the detailed fluctuation relation
given in Eq.~\eqref{eq:ntrdetail1} and the backward probability distribution in Eq.~\eqref{eq:backwardprobantr1}. While the backward probability distribution is positive and normalized, it has no clear operational meaning, i.e., it does not correspond to the measured distribution of an implementable experiment. We stress that the efficacy parameter also loses its operational meaning for measurements without time-reversal symmetry. For completeness, we reprint here the generalized Jarzynski relation following from Eq.~\eqref{eq:ntrdetail1}
\begin{equation}
\label{eq:jarzcg}
\left\langle e^{-\left(\sigma[X,\Lambda(Y)]-\sigma_{\rm cg}[Y]\right)}\right\rangle=1,
\end{equation} 
which implies the second law-like inequality
\begin{equation}
\label{eq:secondlawcg}
\left\langle\sigma[X,\Lambda(Y)]\right\rangle\geq \left\langle\sigma_{\rm cg}[Y]\right\rangle.
\end{equation}
For measurements without time-reversal symmetry, we thus find that our conditions only remedy the shortcomings (i) and (ii) but not (iii). The generalized Jarzynski relation in Eq.~\eqref{eq:jarzcg} has thus the same shortcoming as Eq.~\eqref{eq:jarz2}
in the main text but it results in a strictly more stringent second-law-like inequality.

Alternatively, we can define the backward experiment through the operational meaning of the backward probability distribution in the case where $P_m[Y^\dagger|X^\dagger]= P_m[Y|X]$. This results in Eq.~\eqref{eq:pback}
in the main text which is reprinted here for convenience
\begin{equation}
\label{eq:backwardprobantr2}
P_B[X^\dagger,Y^\dagger]=\frac{P[X^\dagger|\Lambda(Y)^\dagger]}{P[Y^\dagger|\Lambda(Y)^\dagger]}P_m[Y^\dagger|X^\dagger]P[Y].
\end{equation}
As discussed in the main text, this distribution describes an experiment, overcoming shortcoming (iii). We can now relax the condition $P_m[Y^\dagger|X^\dagger]= P_m[Y|X]$ but still keep the backward probability distribution in Eq.~\eqref{eq:backwardprobantr2}. This results in the detailed fluctuation relation
\begin{equation}
\label{eq:ntrdetail2}
\frac{P_B[X^\dagger,Y^\dagger]}{P[X,Y]}=e^{-\left(\sigma[X,\Lambda(Y)]-\sigma_Y-\sigma_m[X,Y]\right)},
\end{equation}
where $\sigma_Y$ denotes the inferable entropy production defined in Eq.~\eqref{eq:detailflucdet}
and we introduced
\begin{equation}
\label{eq:sigmam}
e^{-\sigma_m[X,Y]}\equiv\frac{P_m[Y^\dagger|X^\dagger]}{P_m[Y|X]}.
\end{equation}
Equation \eqref{eq:ntrdetail2} results in the generalized Jarzynski relation
\begin{equation}
\label{eq:jarzsm}
\left\langle e^{-\left(\sigma[X,\Lambda(Y)]-\sigma_Y-\sigma_m[X,Y]\right)}\right\rangle=1,
\end{equation} 
which implies the second law-like inequality
\begin{equation}
\label{eq:secondlawsm}
\left\langle\sigma[X,\Lambda(Y)]\right\rangle\geq \left\langle\sigma_Y\right\rangle+\left\langle\sigma_m[X,Y]\right\rangle.
\end{equation}
We note that the price to pay in order to keep the operational meaning of the backward experiment is that the information term is no longer only dependent on the measurement outcome $Y$. We thus find that shortcomings (ii) and (iii) are overcome by two separate fluctuation relations for measurements without time-reversal symmetry.

\section{E. The Szilard engine}
We consider a particle in a box of volume $v=1$. Starting in thermal equilibrium, the particle is equally likely to be found in the left and in the right half of the box. A partition (wall) is then inserted in the middle of the box and a measurement of the position of the particle is performed. We denote the location of the particle by $x\in \{L,R\}$ and the measurement outcome by $y\in \{l,r\}$. We assume that a measurement error happens with probability $\varepsilon$, i.e.
\begin{equation}
\label{eq:probameascond}
P_m[l|L]=P_m[r|R]=1-\varepsilon,\hspace{2cm}P_m[l|R]=P_m[r|L]=\varepsilon.
\end{equation}
Since the particle is equally likely to be in the left and in the right half of the box, the joint probability for $x$ and $y$ reads
\begin{equation}
\label{eq:probajointszilard}
P[x,y]=\delta_{x,y}(1-\varepsilon)/2+(1-\delta_{x,{y}})\varepsilon/2,
\end{equation}
where the Kronecker delta is defined as $\delta_{L,l}=\delta_{R,r}=1$ and zero otherwise. Having measured $y$, the partition is then moved away from where the particle is assumed to be, extending the volume it presumably occupies to $v_y\leq 1$. 

To evaluate the work extracted in this procedure, we consider the single particle as an ideal gas, described by
\begin{equation}
\label{eq:gaslaw}
k_BT=pv,
\end{equation}
where $p$ is the pressure and $v$ the volume. The extracted work is then given by
\begin{equation}
\label{eq:workgas}
W=\int pdv.
\end{equation}
This results in
\begin{equation}
\label{eq:workszilard}
\beta W[x,y]=\delta_{x,y}\ln(2 v_y)+(1-\delta_{x,{y}})\ln(2-2 v_{y}),
\end{equation}
where $\beta=1/(k_BT)$ denotes the inverse temperature.
The protocol is then completed by removing the partition, such that the particle returns to its initial state. We note that there are two control parameter trajectories, $\Lambda(y)$, which differ by the direction in which the partition is moved upon insertion. In this scenario, the entropy production is determined completely by the work, i.e., $\sigma=-\beta W$.
We note that the work cost diverges if the measurement outcome is erroneous and if $v_{y}=1$ because in this case the particle is squeezed into a vanishingly small volume. For a finite $\varepsilon$ and $v_y=1$, there are thus trajectories for which the entropy production diverges.

We note that because the two control parameter trajectories are the same up to the measurement, the control parameter does not influence the value of $x$ (which is given by the actual particle location when the measurement happens). We therefore find
\begin{equation}
\label{eq:szilardpx}
P[x|\Lambda(y)]=P[x]=\frac{1}{2}.
\end{equation} 
It is then straightforward to verify Eq.~\eqref{eq:probaforward}
in the main text. From Eq.~\eqref{eq:probajointszilard}, we further find that obtaining each measurement outcome is equally likely, i.e., $P[y]=1/2$. The transfer entropy in the forward experiment then reduces to the mutual information
\begin{equation}
\label{eq:mutentszilard}
I[x:y]=\delta_{x,y}\ln(2-2\varepsilon)+(1-\delta_{x,{y}})\ln(2\varepsilon).
\end{equation}
Note that in the limit $\varepsilon\rightarrow 0$, the mutual information diverges when a measurement error occurs because this becomes infinitely unlikely. As a consequence, the standard detailed fluctuation relation involving the mutual information is no longer applicable (see below). Also note that the mutual information does not contain any information on $v_y$. It can thus not take into account any limitation by the protocol we apply. This can be seen most drastically by taking $v_y=1/2$, i.e., the protocol corresponding to doing nothing. Clearly no work can be extracted in this case. The second-law-like inequality involving the mutual information alone does not take this into account [cf.~Eq.~\eqref{eq:jarz1}
]. The mean mutual information reads
\begin{equation}
\label{eq:mutentmeanszilard}
\langle I[x:y]\rangle = \ln(2)+(1-\varepsilon)\ln(1-\varepsilon)+\varepsilon\ln(\varepsilon),
\end{equation}
and is shown in Fig.~\ref{fig:examples}
\,(a). Just as Eq.~\eqref{eq:mutentszilard}, it does not take into account the feedback protocol. Note that the mean mutual information remains finite in the limit of error-free measurements. As noted in Ref.~\cite{sagawa:2012}, the extracted work for a given measurement error is maximized for $v_y=1-\varepsilon$ where $\beta\langle W\rangle =\langle I\rangle$.

The backward experiment discussed in the main text is obtained as follows. First, $\Lambda(y)^\dagger$ is applied with probability $P[y]$. The partition is thus inserted such that the box is divided into parts of volume $v_y$ and $1-v_y$. The partition is then moved to the middle of the box and a measurement of the particle location is performed. The backward experiments are then postselected on the measurement outcomes $y$ which correspond to the applied control parameter (note that in this case $y^\dagger=y$ and $x^\dagger=x$). For the backward experiment, the two control parameter trajectories are different even before the measurement happens. We thus find
\begin{equation}
\label{eq:pxgybackszilard}
P[x|\Lambda(y)^\dagger]=\delta_{x,y} v_y+(1-\delta_{x,{y}})(1-v_y),
\end{equation}
and
\begin{equation}
\label{eq:pygybackszilard}
P[y|\Lambda(y)^\dagger]=\sum_{x=L,R}P_m[y|x]P[x|\Lambda(y)^\dagger]=v_y(1-\varepsilon)+\varepsilon(1-v_y).
\end{equation}
For the joint backward probability distribution we then get from Eq.~\eqref{eq:pback}
\begin{equation}
\label{eq:pbackszilard}
P_B[x,y]=\frac{1}{2}\frac{\delta_{x,y}v_y(1-\varepsilon)+(1-\delta_{x,{y}})\varepsilon(1-v_y)}{v_y(1-\varepsilon)+\varepsilon(1-v_y)},
\end{equation}
and we can easily verify that $\sum_xP_B[x,y]=P[y]=1/2$.

From Eq.~\eqref{eq:detailflucdet}
, we find
\begin{equation}
\label{eq:sigmainfszilard}
e^{-\sigma_y}=2v_y(1-\varepsilon)+2\varepsilon(1-v_y),
\end{equation}
and we can verify the detailed fluctuation relation
\begin{equation}
\label{eq:detailflucszilard}
\frac{P_B[x,y]}{P[x,y]}=e^{\beta W[x,y]+\sigma_y}=\frac{\delta_{x,y} v_y+(1-\delta_{x,{y}})(1-v_y)}{v_y(1-\varepsilon)+\varepsilon(1-v_y)}.
\end{equation} 
We note that for error-free measurements, we obtain $P_B[x,y]=P[x,y]$ and $\beta W=-\sigma_y=\ln(2v_y)$, reflecting the fact that the full entropy production (or extracted work) can be inferred from the measurement outcome $y$. The average of the inferable entropy production is given by
\begin{equation}
\label{eq:avsyszilard}
\langle \sigma_y\rangle =-\ln(2)-\sum_{y=l,r}\frac{1}{2}\ln[v_y(1-\varepsilon)+\varepsilon(1-v_y)].
\end{equation}
Finally, the efficacy parameter is given by
\begin{equation}
\label{eq:gammaszilard}
\gamma=\sum_{y=l,r}P[y|\Lambda(y)^\dagger]=\langle e^{-\sigma_y}\rangle = \sum_{y=l,r}[v_y(1-\varepsilon)+\varepsilon(1-v_y)].
\end{equation}
For $v_l=v_r$, we thus find $\ln(\gamma)=-\langle \sigma_y\rangle$. Otherwise, $\langle \sigma_y\rangle$ gives us a strictly stronger bound on the extracted work. The different bounds on the work obtained by the mutual information, the efficacy parameter, and the inferable entropy are shown in Fig.~\ref{fig:examples}
\,(a).

We close this section with a brief discussion on the conventional definition of the backward probability including feedback
\begin{equation}
\label{eq:backcon}
\tilde{P}_B[x,y]=P[x|\Lambda(y)^\dagger]P[y]=\delta_{x,y} v_y/2+(1-\delta_{x,{y}})(1-v_y)/2.
\end{equation}
This results in the detailed fluctuation relation
\begin{equation}
\label{eq:detailfluccon}
\frac{\tilde{P}_B[x,y]}{P[x,y]}=e^{\beta W[x,y]-I[x:y]}=\delta_{x,y}\frac{v_y}{1-\varepsilon}+(1-\delta_{x,{y}})\frac{1-v_y}{\varepsilon},
\end{equation}
which diverges for $\varepsilon\rightarrow  0$ because $P[x,y]$ is equal to zero for measurement outcomes that do not correspond to $x$ whereas $\tilde{P}_B[x,y]$ remains finite as it is independent of $\varepsilon$.

\section{F. Brownian particle in a harmonic trap}
We consider a Brownian particle in a harmonic trap. Based on the outcome of a position measurement, the minimum of the trap is moved in order to extract work. The particle is initially in thermal equilibrium and the trap potential is centered around $x=0$
\begin{equation}
\label{eq:browninit}
V_0(x)=\frac{k}{2}x^2,\hspace{2cm} P[x]=P[x|\Lambda(y)]=\frac{e^{-\beta V_0(x)}}{\sqrt{2\pi k_BT/k}}.
\end{equation}
As for the Szilard engine, the initial position of the particle, $x$ is independent of the control parameter.
A measurement of position is then performed. We assume the measurement outcome to have a Gaussian distribution
\begin{equation}
\label{eq:pmeasbrown}
P_m[y|x]=\frac{e^{-\frac{(y-x)^2}{2\Sigma^2}}}{\sqrt{2\pi}\Sigma},
\end{equation}
where $\Sigma\rightarrow 0$ corresponds to an error-free measurement. The trapping potential is then shifted such that the minimum coincides with the measurement outcome
\begin{equation}
\label{eq:trappoty}
V_y(x)=\frac{k}{2}(x-y)^2.
\end{equation}
Finally, the system equilibrates in the new trap potential.

The work extracted by this process can be written as
\begin{equation}
\label{eq:workbrown}
W[x,y]=ky(x-y/2),\hspace{2cm}\langle W[x,y]\rangle = \frac{k_BT}{2}- \frac{k\Sigma^2}{2},
\end{equation}
where we used $P[x,y]=P_m[y|x]P[x]$ to evaluate the average. The mutual information (transfer entropy) is given by
\begin{equation}
\label{eq:mutinfbrown}
I[x:y] = \frac{1}{2}\ln\left(\frac{k_BT}{k\Sigma^2}+1\right)-\frac{(x-y)^2}{2\Sigma^2}+\frac{ky^2}{2(k_BT+k\Sigma^2)},
\end{equation}
with an average value of
\begin{equation}
\label{eq:mutavbrown}
\langle I[x:y]\rangle =\frac{1}{2}\ln\left(\frac{k_BT}{k\Sigma^2}+1\right) \geq \beta \langle W[x,y]\rangle,
\end{equation}
where the last inequality can easily be proven. We note that the average mutual information diverges in the error-free measurement limit where $\Sigma\rightarrow 0$. The reason for this is that a perfect position measurement gives an infinite amount of information.

For the backward experiment, the system starts in thermal equilibrium with the external potential $V_y(x)$ chosen with probability
\begin{equation}
\label{eq:pybrown}
P[y]=\int dx P_m[y|x]P[x]=\sqrt{\frac{k}{2\pi(k_BT+k\Sigma^2)}}e^{-\frac{k y^2}{2(k_BT+k\Sigma^2)}}.
\end{equation}
The external potential is then shifted to $V_0(x)$ and the particle location is measured immediately. Finally, the particle thermalizes to recover the initial state. We note that since all variables are position variables, we have $x^\dagger=x$ and $y^\dagger=y$.

The probability that the particle is located at position $x$, given the initial trapping potential $V_y(x)$, reads
\begin{equation}
\label{eq:pxbackcondbrown}
P[x|\Lambda(y)^\dagger]=\frac{e^{-\beta V_y(x)}}{\sqrt{2\pi k_BT/k}}.
\end{equation}
The probability that a position measurement of the particle results in an outcome equal to $y$ reads
\begin{equation}
\label{eq:pybackcondbrown}
P[y|\Lambda(y)^\dagger]=\int dxP_m[y|x]P[x|\Lambda(y)^\dagger]=\sqrt{\frac{k}{2\pi(k_BT+k\Sigma^2)}}.
\end{equation}
From $\gamma = \int dy P[y|\Lambda(y)^\dagger]$, we find that the efficacy parameter diverges. The reason for this is that there are infinitely many control parameter trajectories since there are infinitely many measurement outcomes for a position measurement. The inferable entropy production however remains finite. From Eq.~\eqref{eq:detailflucdet}
in the main text, we find
\begin{equation}
\label{eq:sigmaybrown}
\sigma_Y=-\frac{ky^2}{2(k_BT+k\Sigma^2)},\hspace{1.5cm}\langle \sigma_Y\rangle =-\frac{1}{2}.
\end{equation}
While the efficacy parameter does not provide an inequality, and the mutual information provides an irrelevant inequality as the measurement error becomes small, the inferable entropy production always provides a reasonable bound on the extracted work. The tightness of this bound gives insight into how sharply the measurement resolves the position of the particle.

Finally, from Eq.~\eqref{eq:pback}
in the main text, we find
\begin{equation}
\label{eq:pbackbrown}
P_B[x,y]=\frac{1}{2\pi\Sigma^2}\sqrt{\frac{k\Sigma^2}{k_BT }}\exp{\left[-(x-y)^2\frac{k_BT+k\Sigma^2}{2k_BT\Sigma^2}-\frac{ky^2}{2(k_BT+k\Sigma^2)}\right]},
\end{equation}
and it is straightforward to verify the detailed fluctuation relation
\begin{equation}
\label{eq:flucdetailbrown}
\frac{P_B[x,y]}{P[x,y]}=e^{\sigma_Y+\beta W[x,y]}.
\end{equation}

\end{document}